\documentclass{jfm}

\usepackage{graphicx}
\usepackage[justification=justified]{caption} 
\usepackage{newtxtext}
\usepackage{newtxmath}
\usepackage{natbib}
\usepackage{hyperref}\usepackage{xcolor}
\usepackage{ulem}

\hypersetup{
   colorlinks = true,
   urlcolor   = blue,
   citecolor  = black,
   }

\newcommand{\RomanNumeralCaps}[1]
\linenumbers

\title{From Equilibrium Multistability to Spatiotemporal Chaos in Channel Flows of Nematic Fluids}

\author{Rahil N. Valani\aff{1}
  \corresp{\email{rahil.valani@physics.ox.ac.uk}},
  Sumesh Thampi\aff{1,2}
 \and Julia M. Yeomans\aff{1}}

\affiliation{\aff{1}Rudolf Peierls Centre for Theoretical Physics, Parks Road, University of Oxford, OX1 3PU, United Kingdom
\aff{2}Department of Chemical Engineering, Indian Institute of Technology Madras, Chennai-36, India}

\begin{document}
\maketitle
\begin{abstract}
We investigate channel-confined, nematic liquid crystals using the Beris-Edwards model {of nematohydrodynamics}. Using strong homeotropic anchoring at the walls, we find multistability i.e. multiple coexisting states where the uniform nematic state coexists with state{s} having spatially varying scalar nematic order and director {fields}. When a pressure gradient is applied, flows develop, and the inherent multistability of the system organizes a variety of complex dynamics. For low pressure gradients, steady flows are established, and the director fields that emerge from {the} multistable states at equilibrium correspond to Bowser and Dowser configurations similar to those reported in experiments. An increasing pressure-gradient destabilizes steady Bowser and Dowser flow states sequentially, leading to unsteady periodic and chaotic regimes featuring cyclical topological transitions, pulsating flows, advecting defects and spatiotemporal chaos. These findings demonstrate that modest variations in the scalar nematic order, as captured by the Beris–Edwards model, can qualitatively modify equilibrium structures and give rise to complex nonequilibrium behaviour in confined nematics—contrasting with the Ericksen-Leslie model, which assumes a constant scalar order parameter. Our key model predictions - multistability, periodically oscillating states and advecting defect-mediated turbulence {can} be experimentally investigated in pressure-driven channel flows of nematic fluids.  
\end{abstract}


Broken rotational symmetry and orientationally ordered nematic phases are characteristic of fluids composed of anisotropic constituents, for example molecular liquid crystals \citep{ de1993physics}, colloidal suspensions \citep{nagel2017experimental, lewis2014colloidal} and biological matter \citep{doostmohammadi2022physics, maroudas2021topological}. The continuum theory of nematic liquid crystals provides a fundamental framework for describing the dynamics of such complex fluids \citep{Beris1994, lettinga2005flow, saw2017topological,Hadjifrangiskou}. The microstructure, namely the orientational order of anisotropic constituents can be expressed using a symmetric, traceless tensor, called nematic order-parameter $\mathbf{Q}$. In two-dimensions,
{
\begin{align}
    \mathbf{Q}=
\begin{pmatrix}
Q_{xx} & Q_{xy}\\
Q_{xy} & -\,Q_{xx}
\end{pmatrix}
\equiv S\!\left(\mathbf{n}\mathbf{n}-\tfrac{1}{2}\mathbf{I}\right), \label{eq:qtensor}
\end{align}
}
where the scalar $S$ gives the magnitude of nematic ordering and the vector $\mathbf{n}=(\cos\theta,\sin\theta)$ describes the nematic director, the average orientation of the constituents. Thus, $S$ and $\theta$ capture the two degrees of freedom in the nematic order tensor $\mathbf{Q}$. For isotropic fluids $S = 0$ while for fully ordered nematic fluids $S = 1$. In thermotropic and lyotropic molecular liquid crystals, temperature and concentration respectively dictate the strength of ordering $S$. In soft and biological matter, the additional dominant factors that determine $S$ may be the geometry of the constituents, the nonequilibrium forcing and activity \citep{frenkel2015order, olmsted1992isotropic, Santhosh}.

When driven out of equilibrium either by an external forcing \citep{manneville1981transition, dubois1971hydrodynamic, weiss2013nematic, mur2022continuous} or by an internal activity \citep{giomi2015geometry, alert2022active}, nematic fluids display a wide range of steady and unsteady dynamic behaviours. 
Most theoretical studies that address the dynamics of nematic fluids employ the Ericksen-Leslie equations~\citep{Leslie1968,Stewart2004} which assume a uniform field for the magnitude of the order parameter $S$ and only consider variations in the director field. This simplifies the analysis, and their widespread use may be due to the ease of visualisation of the director field in both experiments and computer simulations.  Exceptions are studies that address the occurrence and dynamics of topological defects. Topological defects represent singularities in the $\mathbf{Q}$ field; the core of the defect has $S = 0$ and an undefined director orientation. Then  a description based on the Beris–Edwards model~\citep{Beris1994}, which incorporates variations in nematic order $S$ is more natural. However, even in the latter theory, in practice, the role of order-parameter variations is often neglected except in the vicinity of the topological defects.


Of particular interest is the case of nematic liquid crystals subjected to pressure-driven flows in confined geometries. Early theoretical work was primarily based on Ericksen-Leslie theory \citep{leslie1987theory}. Further experimental and numerical studies~\citep{ANDERSON201515, Jewell2009, Batista2015_anchoringNematicFlow} have demonstrated that elastic and viscous stresses, together with anchoring conditions at the channel walls, can sustain multiple director configurations, giving rise to flow-driven transitions between distinct nematic states in Poiseuille-type flows. These competing states are often explained using free-energy considerations, with transitions occurring as the pressure gradient shifts the balance between anchoring-dominated and flow-aligned director fields. Subsequent extensions using the Beris–Edwards $\mathbf{Q}$-tensor framework have allowed for systematic treatment of nematic order variations and topological defects, though many analyses continue to neglect order-parameter variations away from defect cores~\citep{Sengupta2013,aplinc2016porous,giomi2017cross}. Recent theoretical studies have further characterised the solution landscapes and bifurcation structure of confined nematic flows, demonstrating how anchoring and confinement geometry mediate the coexistence of Bowser- and Dowser-like states~\citep{Crespo2017_SolutionLandscapesNematicMicrofluidics, Paul2021_AnchoringBifurcationNematicMicroflows}.

Microfluidic experiments using nematic fluids have revealed rich dynamical behaviour consistent with theoretical predictions. Flow-driven transitions between free energy driven and flow-aligned director structures have been observed in nematic channels with strong homeotropic anchoring~\citep{Sengupta2013, Jewell2009}. Recent experiments have demonstrated that careful control of channel geometry, boundary anchoring, and flow rate can stabilise novel intermediate or chiral states~\citep{Copar2020-qu, Ilhan2025-ew}. Classically, Poiseuille-type flows in conjunction with an externally applied magnetic field have been used to manipulate the nematic order and thus determine the rheological properties of nematic liquid crystals \citep{olmsted1992isotropic, kneppe1981comparative}. Together, these studies establish pressure-driven nematic flows as a canonical setting for probing the interplay of rheology, confinement, boundary conditions, and hydrodynamic driving in nematic liquid crystals. However, most analyses have focused on reorienting director fields ($\mathbf{n}$) and their spatial variations, leaving open the role of the magnitude of nematic order ($S$) and its variations in organising the equilibrium and dynamical states in nematic fluids.

In this work, we use the Beris-Edwards $\mathbf{Q}$ tensor formalism and demonstrate that even a slight, spatially uniform reduction in the magnitude of nematic order $S$ can fundamentally alter the equilibrium structure of confined nematic fluids. Specifically, we show that such “global melting” of nematic order gives rise to new equilibrium configurations beyond those predicted by the theories that assume a constant nematic order. When the system is driven out of equilibrium through an imposed pressure gradient, these equilibria act as organizing centres for complex dynamics, including bifurcations between competing configurations, oscillatory states, and spatiotemporal chaos.

The paper is organized as follows: In section~\ref{Sec: theory} we present the theoretical model. In section~\ref{Sec: Eq} we explore the equilibrium state of channel confined nematic fluids (no external forcing). This is followed by analysis of 1D, unidirectional  steady and unsteady flows that nematic fluids develop under external 
forces in sections~\ref{Sec: uni} and \ref{Sec: uni unsteady}, respectively. In section~\ref{Sec: uni unsteady LB} we compare the $1$D results with numerical solutions
of the full $2$D equations. We discuss the implications of the results and conclude in section~\ref{Sec: dis conc}.

\section{Theoretical model}\label{Sec: theory}

We consider an incompressible nematic liquid crystal confined in a $2$D planar channel with the coordinate $-\infty<x<\infty$ along the length of the channel and the coordinate $-L/2\leq y \leq L/2$ along the width of the channel. The Beris-Edwards nematohydrodynamic model~\citep{Beris1994} describes the dynamics of the system in terms of a velocity field $\mathbf{u}=(u_x,u_y)$ and the nematic order parameter field $\mathbf{Q}$ (equation~\ref{eq:qtensor}). 
The governing Beris--Edwards equations read
\begin{align}
\partial_t \mathbf{Q} + \mathbf{u}\!\cdot\!\nabla \mathbf{Q}
&= \mathbf{S}(\nabla\mathbf{u},\mathbf{Q}) + \Gamma\,\mathbf{H}, \label{eq:Qevol}\\
\nabla\!\cdot\!\mathbf{u} &= 0, \label{eq:incomp}\\
\rho\left(\partial_t \mathbf{u} + \mathbf{u}\!\cdot\!\nabla \mathbf{u}\right)
&= \nabla\!\cdot\!\boldsymbol{\sigma}, \label{eq:NS}
\end{align}
where $\rho$ is the mass density and $\Gamma$ is the rotational viscosity.
The strain rate and vorticity tensors are
$\mathbf{E}=\tfrac{1}{2}\!\left(\nabla\mathbf{u}+\nabla\mathbf{u}^{\mathsf{T}}\right)$ and
{$\boldsymbol{\Omega}=\tfrac{1}{2}\!\left(\nabla\mathbf{u}-\nabla\mathbf{u}^{\mathsf{T}}\right)$} respectively, the symmetric and antisymmetric part of the velocity gradient tensor.
Then the generalised co-rotational derivative (flow alignment term) is 
\begin{align}
\mathbf{S}&=(\lambda \mathbf{E}+\boldsymbol{\Omega})\!\cdot\!
\Big(\mathbf{Q}+\tfrac{1}{2}\mathbf{I}\Big)
+\Big(\mathbf{Q}+\tfrac{1}{2}\mathbf{I}\Big)\!\cdot\!(\lambda \mathbf{E}-\boldsymbol{\Omega})-2\lambda\Big(\mathbf{Q}+\tfrac{1}{2}\mathbf{I}\Big)\!\cdot\!\left(\mathbf{Q}\!:\!\nabla\mathbf{u}\right),
\end{align}
with the flow alignment parameter $\lambda$ and the double contraction
$\mathbf{A}\!:\!\mathbf{B}=\mathrm{Tr}(\mathbf{A}^{\mathsf{T}}\mathbf{B})$. The molecular field $\mathbf{H}$ is defined as the variational derivative of  Landau–de Gennes free energy
$\mathcal{F}_\text{nem}=\int f\,\mathrm{d}^2\mathbf{x}$ as
\begin{align}
\mathbf{H} & \equiv -\,\frac{\delta \mathcal{F}_\text{nem}}{\delta \mathbf{Q}}
= -\,\frac{\partial f}{\partial \mathbf{Q}}
+ \nabla\!\cdot\!\frac{\partial f}{\partial (\nabla \mathbf{Q})}.\\ \nonumber
\qquad
\end{align}
We choose
\begin{align}
f&=\frac{C}{2}\!\left(S_{\mathrm{nem}}^{2}-\tfrac{1}{2}\mathbf{Q}\!:\!\mathbf{Q}\right)^{2}
+\frac{K}{2}\, \textcolor{black}{|\nabla \mathbf{Q}|^2}
\end{align}
yielding the explicit form
\begin{equation}\label{eq: H}
\mathbf{H}
= C\!\left(S_{\mathrm{nem}}^{2}-\tfrac{1}{2}\mathbf{Q}\!:\!\mathbf{Q}\right)\mathbf{Q}
+ K\,\nabla^{2}\mathbf{Q}.
\end{equation}
Here $S_{\mathrm{nem}}$ sets the {magnitude of the nematic order parameter} 
which we set to $S_{\mathrm{nem}}=1$,
$C$ controls the stiffness of order parameter variations from $S_{\mathrm{nem}}$, and $K$ is the Frank elastic constant under the single elastic constant approximation.

The total stress field experienced by the nematic fluid combines the Newtonian viscous stress and a back flow stress,
\begin{equation}
\boldsymbol{\sigma} = -p\,\mathbf{I} + 2\eta\,\mathbf{E}
\;\;+\;\boldsymbol{\sigma}_{\mathrm{B}},
\end{equation}
with
\begin{align}
\boldsymbol{\sigma}_{\mathrm{B}} &=
-\lambda\,\mathbf{H}\cdot\Big(\mathbf{Q}+\tfrac{1}{2}\mathbf{I}\Big)
-\lambda\,\Big(\mathbf{Q}+\tfrac{1}{2}\mathbf{I}\Big)\cdot\mathbf{H}\nonumber
+2\lambda\,\Big(\mathbf{Q}+\tfrac{1}{2}\mathbf{I}\Big)\,(\mathbf{Q}:\mathbf{H}) -\frac{\partial f}{\partial (\nabla \mathbf{Q})}:\nabla \mathbf{Q}\nonumber\\
&\quad
+\big(\mathbf{Q}\cdot\mathbf{H}-\mathbf{H}\cdot\mathbf{Q}\big).
\label{stress}
\end{align}


We impose strong homeotropic anchoring boundary conditions at the channel walls $\theta(-L/2)=\theta(L/2)=\pi/2$ with $S(-L/2)=S(L/2)=1$ for the order parameter field, and no-slip, no penetration boundary conditions $\mathbf{u}(-L/2)=\mathbf{u}(L/2)=0$ for the velocity field. We drive the system out of equilibrium by applying a constant pressure-gradient force $F$ along positive $x$ direction. When the unidirectional flow assumption is used in this manuscript, the equations~(\ref{eq:Qevol})--(\ref{stress}) reduce to a $1$D formalism as given in Appendix~\ref{App: oneD}. The resulting partial differential equations~(\ref{Eq: 1D Qvx varyS}) are solved in MATLAB by converting them to ordinary differential equations using the method of lines, with $100$ units of space discretization, and solving the resulting equations using the inbuilt solver ode45. When the fully two-dimensional system is considered, we use a hybrid Lattice-Boltzmann (LB) approach to obtain numerical solutions~\citep{Marenduzzo2007,Thampi2016}. The incompressible Navier-Stokes
equations are simulated in a $100\times20$ domain using a D2Q9 velocity set and a Guo forcing scheme~\citep{Guo2002,Kruger2017}, with periodic boundary conditions along the channel length. {We use lattice Boltzmann units with spatial discretization of  $1$ unit and time step of $1$ unit.} The dynamics of $\mathbf{Q}$ are solved using a Euler–Maruyama finite difference method with a time step of 0.025. Other LB fluid parameters were fixed to: density $\rho=20$, and viscosity $\eta=10/3$. 

\begin{figure}
\centering
\captionsetup{width=\columnwidth}
\includegraphics[width=0.6\columnwidth]{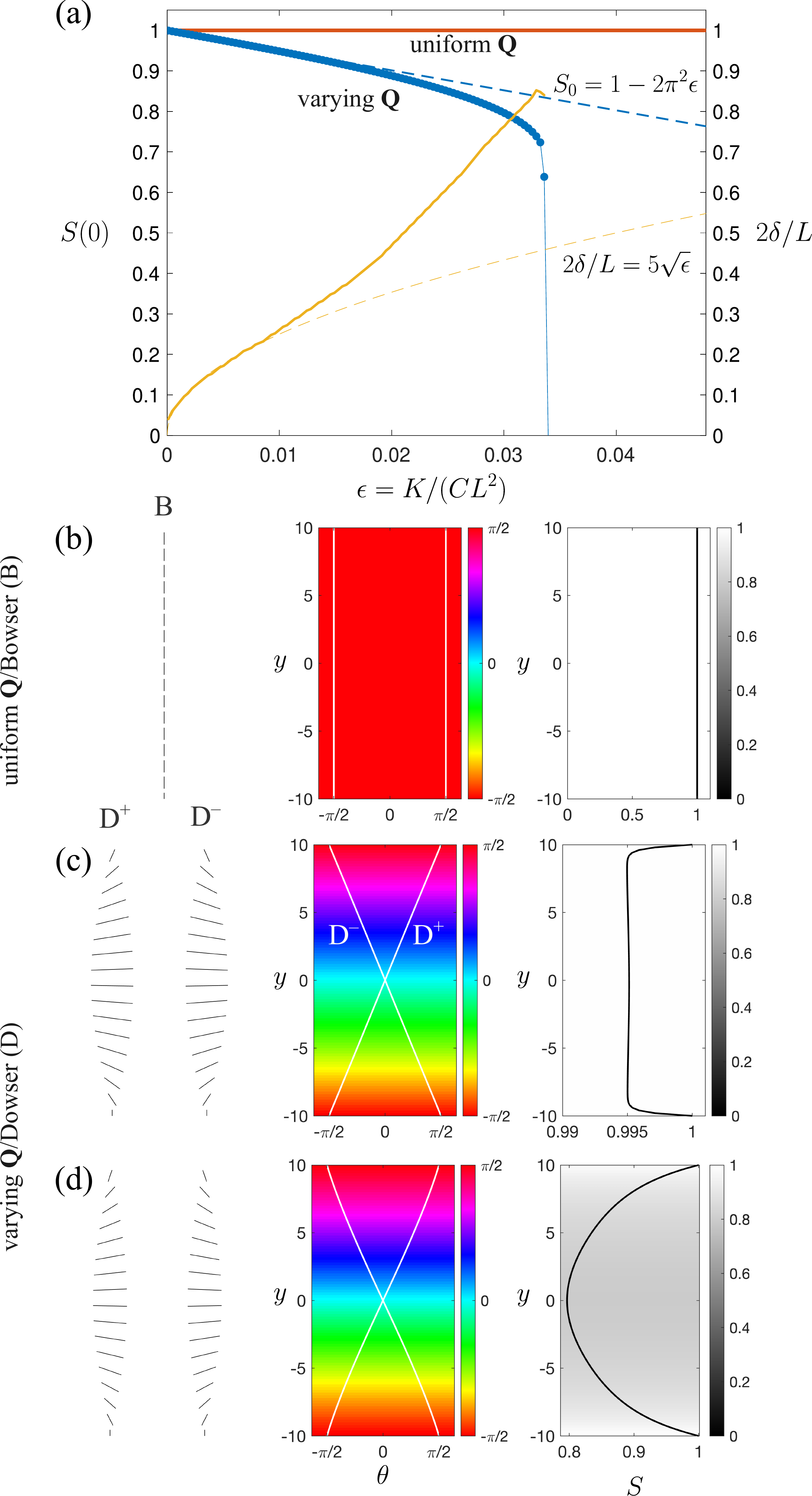}
\caption{Coexistence of {multiple} {equilibrium} states in channel-confined nematics with strong homeotropic anchoring. (a) Nematic order at the centre of the channel $S(0)$ as a function of the dimensionless parameter $\epsilon= \frac{K}{C L^2}$. Simulations were performed by varying $K$ for fixed $C=0.025$ and $L=20$. Red line corresponds to a uniform $\mathbf{Q}$/Bowser (B) state with a constant nematic order of $S(y)=1$. Blue line shows the variation of $S(0)$ in a state with varying $\mathbf{Q}$/Dowser (D). The varying $\mathbf{Q}$ state has a constant nematic order parameter, 
$S_0=1-2\pi^2\epsilon$ (dashed blue line) in the bulk of the channel for small $\epsilon$, whereas strong spatial variations emerge in $S(y)$ for larger $\epsilon$. Yellow curve shows the numerically-calculated boundary layer thickness $\delta$ defined as length from the boundary where nematic order reaches $95\%\,S_0$. Yellow dashed curve shows the fit $2\delta/L=5\sqrt{\epsilon}$. (b) Variations in the scalar nematic order $S$ and nematic orientation $\theta$ for the uniform $\mathbf{Q}$/Bowser (B) state. (c) and (d) show corresponding plots for the varying $\mathbf{Q}$/Dowser (D) states for $\epsilon=0.001$ and $\epsilon=0.03$ respectively.}
\label{Fig: 1}
\end{figure}

\section{{Multiple equilibrium} states in channel-confined nematics}\label{Sec: Eq}


We start by finding the states that minimise the free energy of the channel-confined nematic fluid in the absence of flow (absence of any external forces). Assuming 
translational invariance in the $x$ direction and {using $Q_{xx}=S/2 \cos(2\theta)$, $Q_{xy}= S/2 \sin(2\theta)$}, where $\theta \in (-\pi/2,\pi/2]$ in equations \eqref{eq:Qevol} and \eqref{eq: H}, results in the following ordinary differential equations for the nematic order field and the director field: 

\begin{align}
   0 &= \frac{\text{d} ^2 S}{\text{d} y^2} - 4  S \left(\frac{\text{d} \theta}{\text{d} y} \right)^2 + \frac{C}{K} S (1 - S^2), \label{Eq: 1D Sthetavx}\\
  0 &=   \frac{\text{d}^2 \theta}{\text{d} y^2} + \frac{2 }{S} \frac{\text{d} \theta}{\text{d} y} \frac{\text{d} S}{\text{d} y}. \label{Eq: 1D Sthetavx2}
\end{align}
Numerically solving equations~\eqref{Eq: 1D Sthetavx}-\eqref{Eq: 1D Sthetavx2}, with strong homeotropic anchoring at the channel boundaries, we find multiple coexisting solutions. 

As expected, the solution with uniform $\mathbf{Q}$, i.e. $S(y)=1$ and $\theta(y)=\pi/2$ (see figure~\ref{Fig: 1}(b)), corresponds to the global minimum of the nematic free energy $\mathcal{F}_{\mathrm{nem}}$. However, there are also other solutions - where $\mathbf{Q}$ varies across the channel~(see figure~\ref{Fig: 1}(c)). These correspond to local minima of the free energy since spatial variations in $S$ and $\theta$ result in positive free energy contributions. They are topologically different to the uniform $\mathbf{Q}$ state in that they have $\theta=0$ at the centre of the channel, and an overall change in the nematic orientation from one wall to the other of $n\pi$. {We shall consider the single-twist states with $n=1$ in the remainder of this paper.} We will refer to the uniform $\mathbf{Q}$ state as Bowser (B) and the $n=1$ varying $\mathbf{Q}$ states as Dowser (D), since these steady states will give rise to the well known Bowser and Dowser states~\citep{Copar2020-qu}  in pressure-driven flows (see section~\ref{Sec: uni}).


The Dowser states exist as degenerate pairs, distinguished by their chirality, D$^+$ and D$^-$. For these states the orientation (the director field) varies almost linearly across the channel $\theta(y)\approx\pm\pi y/L$. The spatial variation in the nematic order $S$ {can have} two qualitatively different forms:
\begin{enumerate}
    \item $S(y) = $constant but slightly less than unity in the bulk of the channel (figure~\ref{Fig: 1}(c))
    \item $S(y)$ has a smooth variation (a parabolic shape) across the channel (figure~\ref{Fig: 1}(d)).
\end{enumerate}
To quantify the crossover between these forms of $S(y)$ of the Dowser states 
we define a dimensionless parameter $\epsilon=K/(CL^2)$ that compares the nematic correlation length ($\sqrt{K/C}$) with the width of the channel (L). Figure~\ref{Fig: 1}(a) shows the value of the scalar nematic order parameter at the centre of the channel $S(0)$ as a function of $\epsilon$. Small $\epsilon$ corresponds to spatial variations in $S$ of the form (i) whereas larger values of $\epsilon$ correspond to $S(y)$ of form (ii). 

For the Dowser states at small $\epsilon$, 
{the value of the nematic order parameter} can be approximated from Eqs.~\eqref{Eq: 1D Sthetavx} assuming a constant $S=S_0$ and a linearly varying director field. This results in a balance of {contributions} from free energies corresponding to nematic order and gradients in director giving
\begin{align*}
4  S \left(\frac{\partial \theta}{\partial y} \right)^2 \approx \frac{C}{K} S (1 - S^2).
\end{align*}
Using $\partial \theta/\partial y \approx \pi/L$ and $S\approx S_0$ we get
\begin{equation}
    S_0 \approx 1 - \frac{2 K \pi^2}{C L^2} = 1 - 2\pi^2 \epsilon.  
\end{equation}
This equation (plotted as a dashed blue line in figure~\ref{Fig: 1}(a)) shows  that without any spatial variations in the nematic order $S$ in the channel bulk, a small, uniform decrease in the magnitude of the nematic order {($\approx 2\pi^2 \epsilon$)} can support spatial (linear) variations in the orientation of the nematic field{, and maintain the state of equilibrium}. 

We can further analyse the boundary layers that are formed for $S$ near the channel walls for small $\epsilon$. By non-dimensionalizing the length scale with $L/2$ and performing a boundary layer analysis (see Appendix~\ref{App: boundary layer}) we obtain an asymptotic solution for the nematic order variations within the channel,
$$S(y)=1-2\pi^2\epsilon + 2\pi^2\epsilon \left( e^{-\frac{1-y}{\sqrt{2\epsilon}}} + e^{-\frac{1+y}{\sqrt{2\epsilon}}} \right) + O(\epsilon^2).$$
The boundary layer thickness $\delta \propto \sqrt{\epsilon}$. Figure~\ref{Fig: 1}(a) shows a comparison of this scaling (yellow dashed curve) with value calculated numerically (yellow solid curve).

\section{Unidirectional steady flows}\label{Sec: uni}


\begin{figure}
\centering
\captionsetup{width=\columnwidth}
\includegraphics[width=\columnwidth]{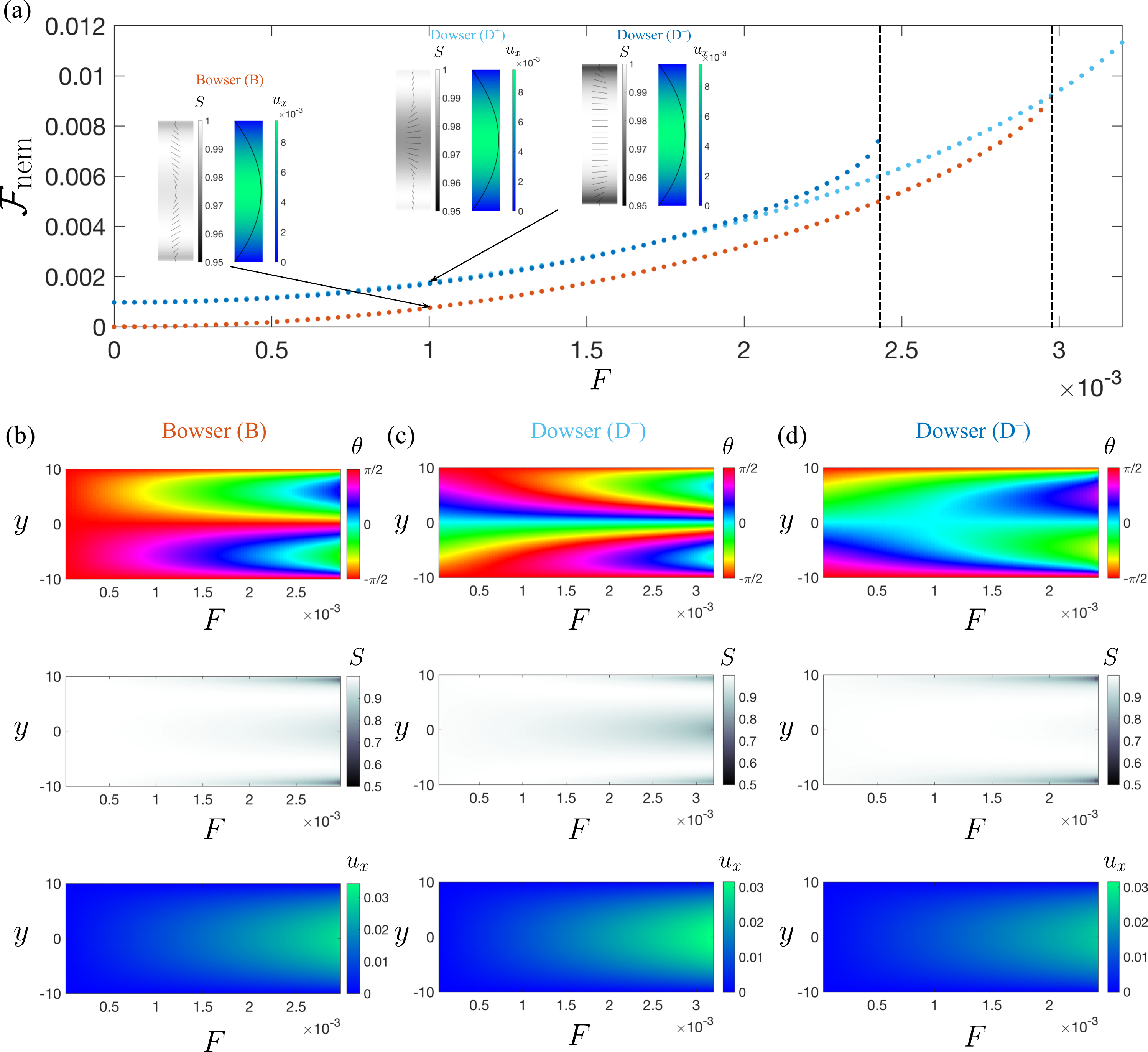}
\caption{{Nonequilibrium} steady flow states for pressure-driven nematics in a channel. Panel (a) shows the nematic free energy $\mathcal{F}_{\mathrm{nem}}$ as a function of the applied pressure-gradient force $F$ for B (red dots), D$^-$ (blue dots) and D$^+$ (cyan dots) states. (b)-(d) show the evolution of (top) nematic orientation $\theta$, (middle) scalar nematic order $S$ and (bottom) velocity $u_x$ for B, D$^+$, and D$^-$ states, respectively. Bowser states originate from a uniform $\mathbf{Q}$ state at equilibrium ($F=0$) whereas Dowser states originate from varying $\mathbf{Q}$ states. The inset in panel (a) shows example nematic configuration ($S(y)$ and $\theta(y)$) and velocity field ($u_x(y)$) at $F=0.001$ for Bowser and Dowser states. Vertical dashed lines in (a) show the location of instability of the D$^-$ and B states. Other parameters were fixed to $C=0.025$, $K=0.01$, $L=20$, $\lambda=0$.}
\label{Fig: 2_1}
\end{figure}

The Bowser (B) and Dowser (D$^\pm$) states that correspond to free energy minima of channel confined nematics can be driven out of equilibrium by applying a pressure-gradient, $F$, along the length of the channel. We investigate the variations of the steady states with $F$ by numerically solving the $1$D unidirectional flow equations~\eqref{Eq: 1D Qvx varyS}. In figure~\ref{Fig: 2_1}(a) we plot a bifurcation diagram that shows how the global free energy of each state varies with $F$.

The forcing induces a Poiseuille-like unidirectional velocity field in the channel. As a result the director orientation of each of the states begins to deviate from its equilibrium configuration, as shown in the inset of figure~\ref{Fig: 2_1}(a), and hence the nematic free energy increases.
Importantly, we find that the steady states become unstable, one by one, with increasing $F$. For the parameter combinations shown in figure~\ref{Fig: 2_1}, the D$^-$ state becomes unstable first, followed by the instability of the B state and then, eventually, the instability of D$^+$ state. In general, the order of instability of the states depends on the system parameters and a different sequence of instabilities is found for different combinations of parameters~(see Appendix~\ref{App: 1}).




{Contour plots} showing variations in the nematic director, nematic order and the unidirectional velocity field as a function of $F$ for the B, D$^+$, and D$^-$ states, respectively, are shown in figure~\ref{Fig: 2_1}(b)-(d). As $F$ is progressively increased, there is melting of the nematic order parameter $S$ near the walls for the D$^-$ state, whereas for the B and D$^+$ states, we find a reduction in the nematic order both near the channel centre and near the channel walls. Furthermore, we see significant distortions of the nematic director field with increasing $F$ in all three states, concentrated in the regions of lowest nematic ordering.

The existence of Bowser and Dowser states has been reported in experiments with pressure-driven nematics where typically a transition from a Bowser to a Dowser state is observed with increasing flow rate in $3$D channel flows~\citep{Sengupta2013,Copar2020-qu,Ilhan2025-ew,Jewell2009}. Attempts have been made in the literature to rationalize the Bowser-Dowser transitions using the Erickson-Leslie equations that assume variations in only the nematic orientation~\citep{ANDERSON201515,Jewell2009,Batista2015_anchoringNematicFlow,Crespo2017_SolutionLandscapesNematicMicrofluidics,Paul2021_AnchoringBifurcationNematicMicroflows}. 
In the Ericksen-Leslie formalism, the two topologically distinct Bowser and Dowser flowing steady states cannot be simultaneously realized using different initial conditions and one typically realizes them via different anchoring boundary conditions at the channel walls. In this formalism, transition between the two states is usually interpreted using free energy arguments where a Bowser-Dowser transition occurs because the  Dowser state has a lower free energy at larger applied pressure-gradient forces $F$. The $\mathbf{Q}$-tensor formalism presented here, that takes into account both variations in nematic order and the nematic director, can support multistability i.e. coexisting Bowser and Dowser states. This suggests that the transitions between Bowser and Dowser states that are observed in experiments might be due to a dynamical bifurcation i.e. one of the states becoming unstable and the system converging onto the remaining stable branch, as opposed to the system trying to minimize the nematic free energy. 

However, we note that since the experiments are mostly performed with $3$D channels, the specific sequence and details of transitions that are observed might be dependent on the $3$D nature of the geometry.
Nevertheless, the multistability of Bowser and Dowser states can be probed in experiments via hysteresis effects by, for example, starting in a Bowser state and slowly increasing the pressure gradient/flow rate up to the point of a Bowser-Dowser bifurcation, and then slowly decreasing the pressure gradient/flow rate. If hysteresis is indeed observed in experiments, then it implies coexistence of Bowser-Dowser states in the underlying system.





\begin{figure}
\centering
\captionsetup{width=\columnwidth}
\includegraphics[width=\columnwidth]{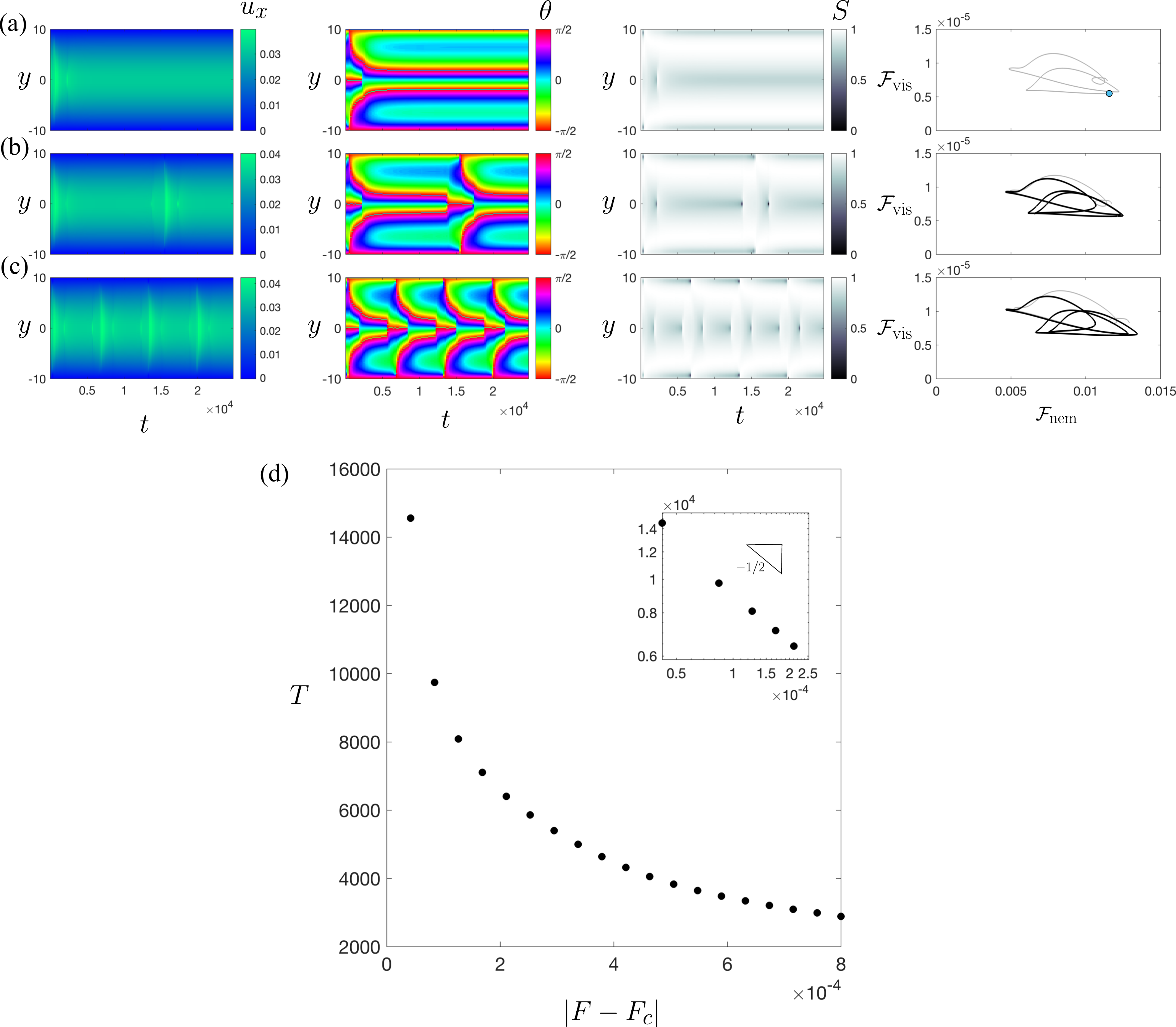}
\caption{Transition from steady to unsteady oscillatory states. Panels (a)-(c) show, from left to right, kymographs of (i) velocity field $u_x$, (ii) nematic orientation $\theta$, (iii) scalar nematic order $S$, and (iv) phase-space projection of the dynamics onto the $2$D space of global viscous dissipation $\mathcal{F}_{\mathrm{vis}}$ and global free energy $\mathcal{F}_{\mathrm{nem}}$ (gray curve show transients whereas black curve show the stable attractor). (a) corresponds to $F=0.0032$ where the system relaxes to the D$^+$ steady state (cyan circle), whereas (b) and (c) correspond to $F=0.00324$ and $F=0.0034$, respectively, after the onset of oscillations. Panel (d) plots the period $T$ of oscillations as a function of proximity to the transition, and from the inset a power law scaling for the divergence of the form $T\propto |F-F_c|^{-1/2}$ is evident. Other parameters of the system are fixed to $C=0.025, K=0.01$, $L=20$ and $\lambda=0$.}
\label{Fig: 3}
\end{figure}


\section{Unsteady unidirectional flows}\label{Sec: uni unsteady}

\begin{figure}
\centering
\captionsetup{width=\columnwidth}
\includegraphics[width=\columnwidth]{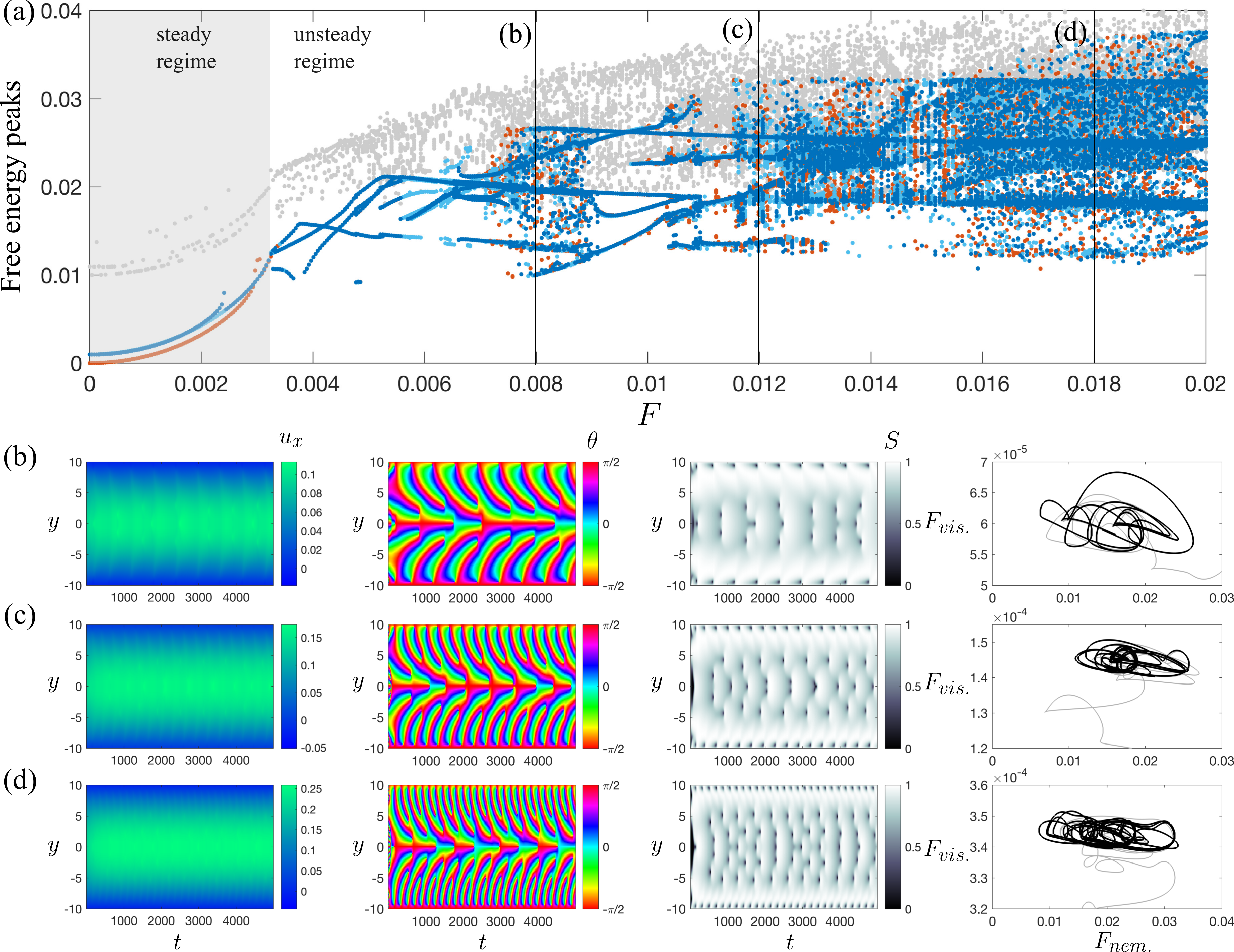}
\caption{Unsteady dynamics in pressure-driven nematic flows. (a) Bifurcation diagram showing peaks in the time series of the global free energy as a function of $F$ for unidirectional, pressure-driven flow setup. Initial flow velocity is zero, $S=1$, and different colours correspond to different initial orientation profiles with red favouring B states, blue and cyan favouring D$^-$ and D$^+$ states, respectively, and grey corresponding to random initial orientations. Panels (b)-(d) show kymographs of velocity, nematic orientation, scalar nematic order and phases-space trajectories (transients in grey and stable attractors in black) for $F=0.008, 0.012$ and $0.018$, respectively. Other parameters of the system are fixed to $ C=0.025, K=0.01$, $L=20$ and $\lambda=0$.}
\label{Fig: 5}
\end{figure}

At large values of $F$, beyond those shown in figure~\ref{Fig: 2_1}(a), all steady flows states become unstable and the system enters an unsteady dynamical regime. At the onset of this regime, we find oscillatory states of the system. The transition from a steady state to an oscillatory state is illustrated in figure~\ref{Fig: 3}(a)-(c) for increasing values of $F$. These plots show the kymograph of the unidirectional velocity field $u_x(y)$, nematic orientation field $\theta(y)$ and the nematic order field $S(y)$, together with a phase-space projection in the space of the global nematic free energy $\mathcal{F}_{\mathrm{nem}}$ and the global viscous dissipation associated with the fluid flow $\mathcal{F}_{\mathrm{vis}}$, defined as
$$\mathcal{F}_{\mathrm{vis}}=\int_{-L/2}^{L/2} \frac{\eta}{2}\left(\frac{1}{2}\frac{\partial u_x}{\partial y}\right)^2\, dy.$$

The oscillating state corresponds to cyclical transitions between (unstable) Bowser-like and Dowser-like states. Since these states are topologically distinct, the periodic transitions between them are mediated by topological transitions i.e. a {continuous line resembling a} line of defects in the $2$D channel. This can be inferred from the kymograph of the nematic orientation and the nematic order shown in figure~\ref{Fig: 3}(b)-(c) where the nematic order parameter $S$ goes to zero. Note, that {$S$ goes to zero at} the centre or near the walls of the channel. The oscillatory nature of the flow emerges from the two-way coupling of the nematic field and the velocity field in the nematohydrodynamic model. 

We observe oscillating states with very large period T at the onset of the transition and the oscillation period decreases progressively with increasing $F$ as shown in figure~\ref{Fig: 3}(d). Furthermore, we obtain a power-law scaling behaviour (see inset) for the divergence of the oscillation period as a function of the proximity to the transition of the form $T\propto |F-F_c|^{-1/2}$ where $F_c\approx 0.00321$ is the critical value of $F$ where the bifurcation from steady flow to oscillatory flow occurs. A critical exponent of $-1/2$ is a characteristic of saddle-node type bifurcations~\citep{Strogatz2015_NonlinearDynamicsChaos} and suggests that the onset of oscillations take place via an infinite-period bifurcation as opposed to a Hopf bifurcation that is ubiquitously associated with the onset of oscillations for many nonlinear dynamical systems. This is further evident from the phase space trajectory in figure~\ref{Fig: 3}(a) just before the bifurcation where the trajectory performs a large excursion, almost traversing the limit cycle, before converging onto the D$^+$ stable point (cyan circle). This suggests that the D$^+$ fixed point loses stability in a saddle-node bifurcation and the resulting emergence of oscillations is via a saddle-node, infinite-period (SNIPER) bifurcation. 

To characterize the full sequence of transitions with increasing $F$, we plot a bifurcation diagram  spanning both the steady and the unsteady regimes  in figure~\ref{Fig: 5}(a). The quantity we choose to plot in the bifurcation diagram are the peak values in the time series of global nematic free energy i.e. $\mathcal{F}_{\mathrm{nem}}(t_n)$ where $t_n$ are time stamps corresponding to a local maxima in the free energy.  In the steady regime, since $\mathcal{F}_{\mathrm{nem}}$ is a constant, this would correspond to a single point on the bifurcation diagram. In the unsteady regime, a finite set of points correspond to periodic motion whereas scattered points typically correspond to chaotic or quasiperiodic dynamics.

To detect and characterize the presence of multistability i.e. the existence of different states at the same parameter values, the bifurcation diagram is created by solving the one-dimensional equations of motion~(\ref{Eq: 1D Qvx varyS}) with different initial profiles for the nematic orientation field $\theta(y,t=0)$. We choose:
\begin{align} \label{initialcond}
\theta_{B}(y,t=0)&=\frac{\pi}{2}+\frac{F}{12K(\eta\Gamma+1)} y \left(\left(\frac{L}{2}\right)^2-y^2\right),\\ \nonumber
\theta_{D+}(y,t=0)&=\frac{\pi}{L}y+\frac{F}{12K(\eta\Gamma+1)} y \left(\left(\frac{L}{2}\right)^2-y^2\right),\\ \nonumber
\theta_{D-}(y,t=0)&=-\frac{\pi}{L}y+\frac{F}{12K(\eta\Gamma+1)} y \left(\left(\frac{L}{2}\right)^2-y^2\right),\\ \nonumber
\theta_{R}(y,t=0)&=\left[\frac{-\pi}{2},\frac{\pi}{2}\right]+\frac{F}{12K(\eta\Gamma+1)} y \left(\left(\frac{L}{2}\right)^2-y^2\right). 
\end{align}

The first term in these expressions corresponds to the orientation of the B, D$^+$ and D$^-$ states at equilibrium, or R, a random initial orientation where $\theta_R(y,0)$ is uniformly sampled between $[\frac{-\pi}{2},\frac{\pi}{2}]$, respectively. The second term accounts for the distortions in the nematic orientation due to an applied Poiseuille flow velocity profile for a constant scalar nematic order field. The initial nematic order field is set to $S(y,t=0)=1$ and the unidirectional velocity field is set to $u_x(y,t=0)=0$. 

The solutions are shown in  figure~\ref{Fig: 5}(a) as red for states favouring Bowser configurations $\theta_B(y,0)$, cyan for Dowser +, $\theta_{D+}(y,0)$, dark blue for Dowser -, $\theta_{D-}(y,0)$ and gray for random initial orientations, $\theta_{R}(y,0)$. The steady regime of the bifurcation diagram is similar to figure~\ref{Fig: 2_1}(a) for initial conditions favouring Bowser and Dowser states, whereas random initial conditions take the system to higher order Dowser states with $n>1$. (We note that in figure~\ref{Fig: 2_1}(a), we used numerical continuation to follow the steady branches when they are stable, whereas here we solve the PDEs with the 
initial conditions specified above.) Once all the steady states (gray region) become unstable, we first see oscillatory unsteady behaviour from $F\approx 0.0034$ to $F\approx 0.007$, with chaotic behaviour first emerging beyond $F\approx 0.007$, for initial conditions favouring Bowser and Dowser states.  By contrast, for random initial conditions, the scattered dots indicate chaos for almost all values of $F$ in the unsteady regime. We also notice from the bifurcation diagram that, if the initial conditions are not random, the system does not permanently maintain chaos for large $F$ in the unsteady regime, and we find alternating regimes of periodic and chaotic behaviours with increasing $F$; a feature that is common in bifurcation diagrams of low-dimensional chaotic systems~\citep{Strogatz2015_NonlinearDynamicsChaos}. 

Kymographs of the chaotic dynamics with increasing $F$ beyond $F\approx 0.007$ are displayed in figure~\ref{Fig: 5}(b)-(d). These show that, although the dynamical behaviour is not entirely periodic, its chaotic nature is weak. There is periodicity on short time scales, with period decreasing with $F$, and irregularities emerging at longer time scales spanning several periods. This is further evident from the phase space trajectories which show the system traversing a chaotic attractor with some structure, a signature of low-dimensional chaos \citep{Strogatz2015_NonlinearDynamicsChaos}. The almost periodic nature of the chaos is also evident from the bifurcation diagram where, in the chaotic regime, the points are not scattered uniformly but are concentrated near periodic branches. These results provide evidence that, beyond the periodic regime, the system transitions to weak spatiotemporal chaos. Thus, a unidirectional, pressure-driven nematic exhibits a variety of dynamical regimes: steady flows, unsteady periodic flows and spatiotemporal chaos (see Appendix~\ref{App: 1} for bifurcation diagrams for other parameter values). 
~\\

\begin{figure}
\centering
\captionsetup{width=\columnwidth}
\includegraphics[width=\columnwidth]{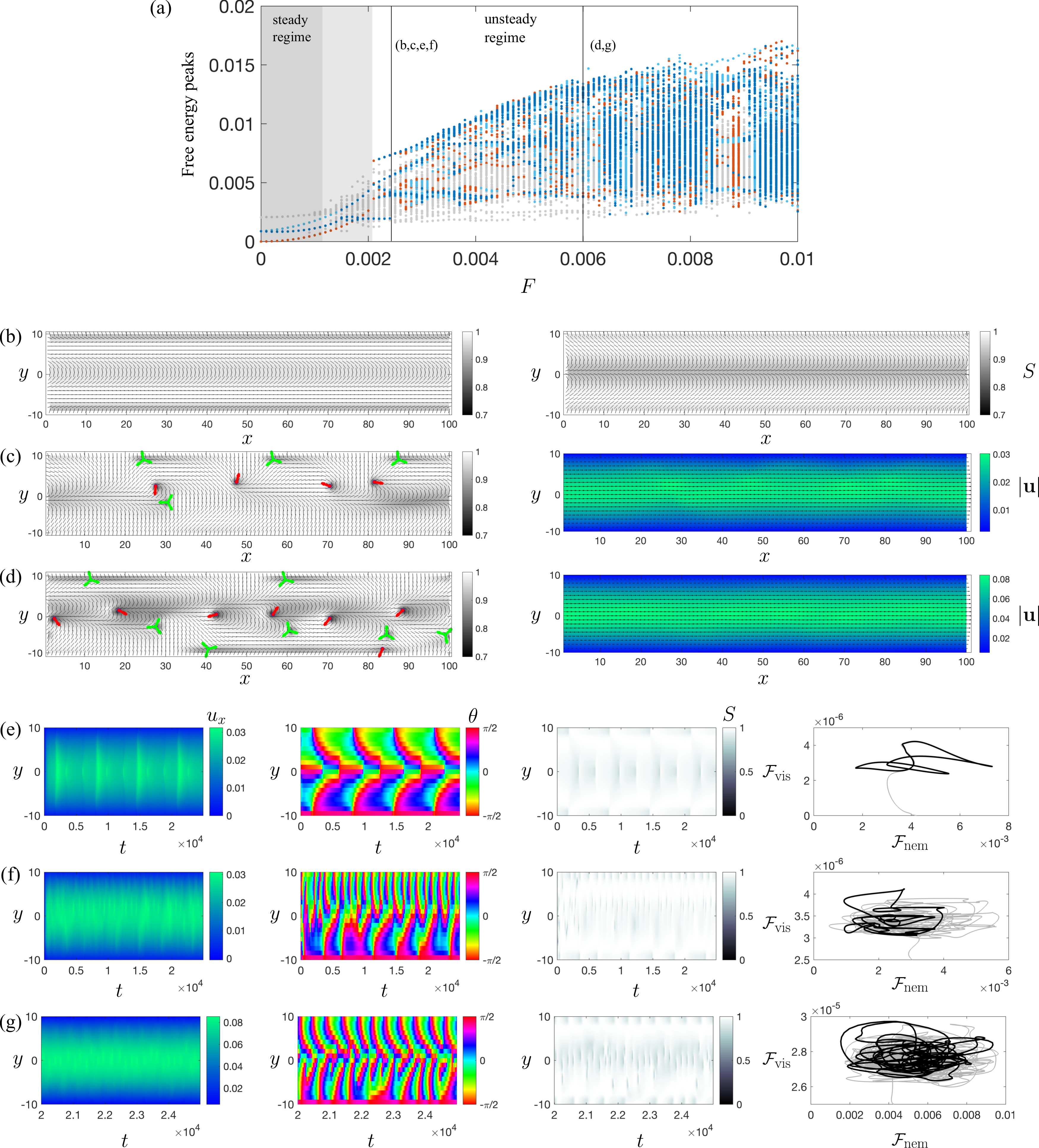}
\caption{Nematohydrodynamics in $2$D pressure-driven channel flows. (a) Bifurcation diagram analogous to figure~\ref{Fig: 5}(a) for the full $2$D equations of motion~~(\ref{eq:Qevol})--(\ref{eq:NS}). In $2$D flows, multistability is observed between different unsteady dynamical states. For example, at $F=0.0024$, (b,e) spatial uniform initial conditions along $x$ corresponding to a Bowser or a Dowser configuration result in periodic states where the whole channel periodically transitions between a Bowser-like and a Dowser-like state. Conversely, if the initial conditions are random, (c,f) then defects are stabilised by advection along the channel. This results in weakly chaotic flow, as the defect dynamics are not entirely periodic. (d,g) At a higher value of $F=0.006$, the turbulent flow becomes more prominent through the continuous formation, annihilation and advection of defects. The kymographs in (e)-(g) are shown for at a fixed $x=50$. Other parameter values are the same as for figure~\ref{Fig: 5}.}
\label{Fig: 6}
\end{figure}

\section{Unsteady flows in $2$D channel} \label{Sec: uni unsteady LB}

We now solve the full 2D nematohydrodynamic equations of motion~(\ref{eq:Qevol})--(\ref{stress}) using a hybrid lattice Boltzmann algorithm~\citep{Marenduzzo2007,Thampi2016}. 

Analogous to figure~\ref{Fig: 5}(a), we consider pressure-driven $2$D channel flows and create a bifurcation diagram as a function of $F$, as shown in figure~\ref{Fig: 6}(a). To most easily compare the flow dynamics between $2$D and $1$D, we choose the bifurcation parameters as the peak values in the time series of the nematic free energy calculated at a fixed $x$ location along the channel and integrated across the width $y$ of the channel. The initial conditions are chosen to be the same as those for the bifurcation diagram in figure~\ref{Fig: 5}(a), equation~(\ref{initialcond}), where we assume the initial orientation field to vary in the $y$ direction but to be constant in the $x$ direction for the Bowser and Dowser favouring initial configurations.

For small values of $F$, we find Bowser and Dowser steady flowing states (grey region) and they undergo the same sequence of instabilities as observed in the $1$D formalism.
{Further,} we ubiquitously find multistability between steady and unsteady states, as well as between different types of unsteady states. In particular, for $F\sim0.0015$ (light grey region in figure~\ref{Fig: 6}(a)), the steady states coexist with a dynamic advecting defect state. Initial conditions that favour Bowser or Dowser states indeed lead to Bowser or Dowser steady states with variations in orientation only in the $y$ direction as for the $1$D model. However, if the initial conditions are random, there are many $+1/2$ and $-1/2$ topological defects~\citep{Doostmohammadi2018_ActiveNematics} present in the initial state. Not all of these are able to annihilate with each other and they are advected down the channel by the applied pressure gradient. 

With increasing $F$, there is multistability in unsteady states {as well}. 
For example, for $F=0.0024$, Bowser-like or Dowser-like initial conditions lead to the entire $2$D channel undergoing uniform periodic oscillations between a Bowser and a Dowser state (figure~\ref{Fig: 6}(b)). The kymographs of the velocity field, nematic orientation and nematic order and the phase-space trajectory, which are plotted for a fixed $x$ in figure~\ref{Fig: 6}(e),  are similar to those obtained in $1$D (compare figure~\ref{Fig: 3}(a)-(c)). 
{Instead,} if the system is started with random initial conditions, we obtain advecting defects, showing signs of weak chaos, as shown by the snapshot in figure~\ref{Fig: 6}(c). The corresponding kymographs are shown in figure~\ref{Fig: 6}(f). 

Further increases in $F$ lead to more irregular motion of defects i.e. spatiotemporal chaos or topological turbulence, where defects are created and annihilated as they advect down the channel as shown in figure~\ref{Fig: 6}(d), with the corresponding kymographs in figure~\ref{Fig: 6}(g). 
Note that $+1/2$ defects are typically found near the centre of the channel whereas $-1/2$ defects are found near the channel walls.
 This is consistent with the turbulent state observed in the $1$D model and also with variations in the nematic order observed in steady states for large $F$ (see figure~\ref{Fig: 2_1}(a)), in both cases the nematic order melts most strongly near the channel centre or walls. 

\begin{figure}
\centering
\captionsetup{width=\columnwidth}
\includegraphics[width=0.75\columnwidth]{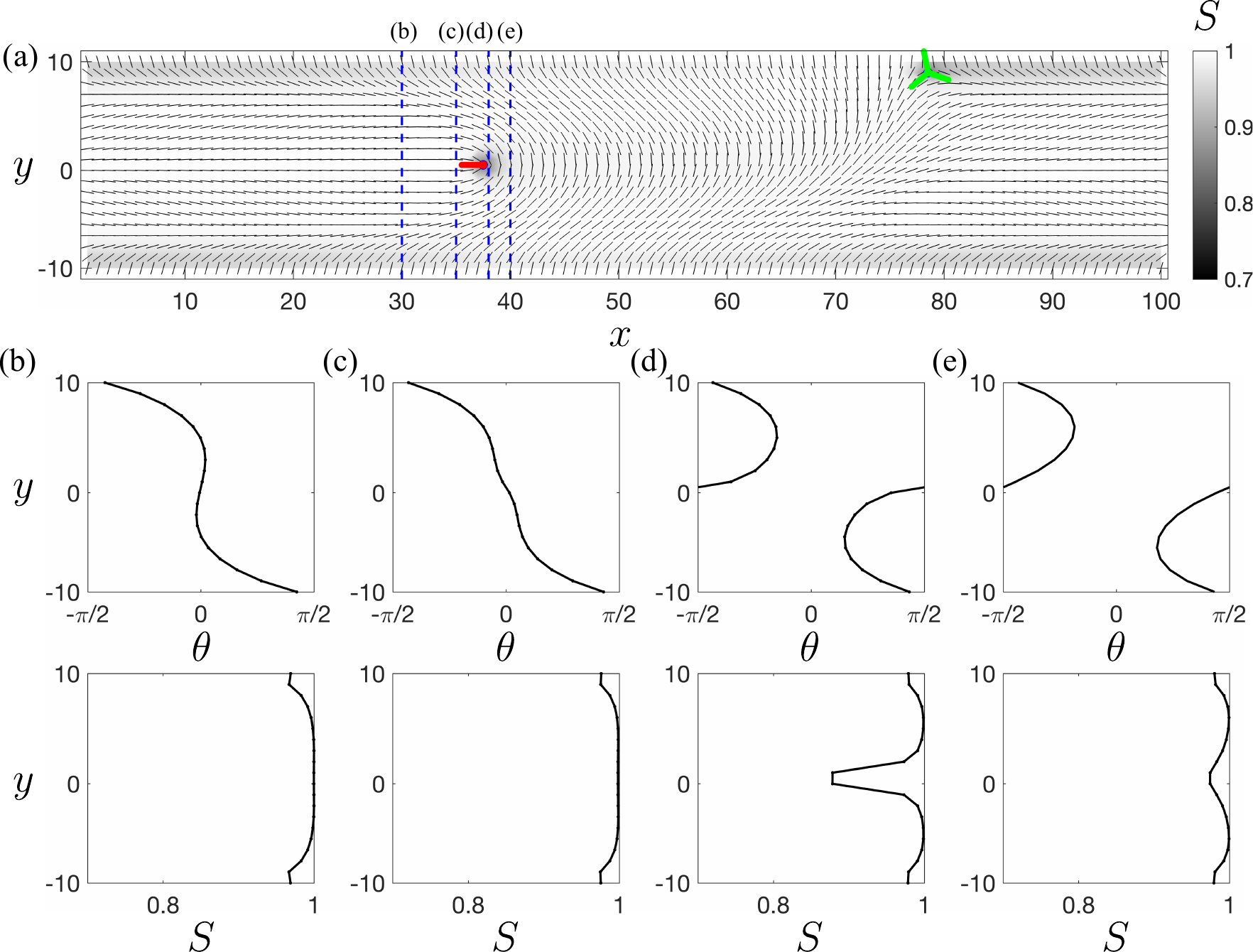}
\caption{Topological defects mediate spatially localized Bowser-Dowser transitions in $2$D pressure-driven nematic flows. (a) Snapshot from a simulation at $F=0.014$ with random initial conditions showing defects advecting along the channel. The  $1$D profiles at different locations along the channel near a defect are shown in panels (b)-(e).  A Bowser-Dowser transition is obtained in the $1$D profile as the defect is traversed spatially.}
\label{Fig: 7}
\end{figure}

The essence of spatiotemporal chaos in $2$D driven nematics that we report here lies in the irregular dynamics of topological defects as they are created, annihilated and advected down the channel due to an applied strong pressure-gradient $F$. The creation and annihilation of the defects is a reflection of local topological transition between Bowser-like and Dowser-like states. To show this more clearly, we consider an advecting defect state at $F=0.0014$ obtained from random initial conditions. After initial transients, we obtain advecting pairs of defects as shown in figure~\ref{Fig: 7}(a). Plotting the nematic order and nematic orientation fields at different $x$ positions in the vicinity of the $+1/2$ defect shows a transition from a Bowser-like to a Dowser-like state (figure~\ref{Fig: 7}(b)-(d)). Hence, topological defects facilitate spatially localized Bowser-Dowser transitions in nematic channel flows. 



\section{Discussion and conclusions}\label{Sec: dis conc}

We investigated channel flows of nematic fluids using the Beris–Edwards nematohydrodynamics formalism, comparing a one-dimensional unidirectional model to solutions of the fully two-dimensional differential equations. For a channel confined nematic fluid, at equilibrium (in the absence of an applied pressure gradient), we found coexisting states of the nematic: in addition to the expected solution of uniform $\mathbf{Q}$, we found that solutions of varying $\mathbf{Q}$ across the width of the channel are also stable. These states are distinct, they have different topology characterized by a difference in winding number of the director when calculated from the bottom wall to the top wall of the channel. When the system is driven out of equilibrium by an imposed pressure gradient, these equilibrium states serve as reference points, about which we observe  bifurcations between competing configurations, oscillatory behaviour, and spatiotemporal chaos.


At small applied pressure gradients, we obtain a steady Poiseuille-like unidirectional velocity field. Correspondingly, the equilibrium nematic configurations develop distortions. The resulting director fields are the well known Bowser and Dowser states for channel flows of nematic liquid crystals. Increasing pressure gradients lead to the instability of these steady flowing states, resulting in transitions between Bowser and Dowser states. Bowser-to-Dowser transitions have been observed in three-dimensional microfluidic nematic channel flows and are commonly explained in the literature using free-energy arguments. We show instead that both Bowser and Dowser states originate from coexisting, topologically distinct equilibrium configurations, and that transitions between them can be interpreted as dynamical instabilities arising when the steady states lose stability.

At larger pressure-gradients, we found that all the steady states cease to exist, and oscillations emerge in both the nematic field and the velocity field. These periodic oscillations correspond to cyclical transitions between Bowser-like and Dowser-like states. 
The period is very large at the onset of oscillations, and it decreases with increasing pressure gradient. Near the bifurcation, the decrease follows a $-1/2$ power law consistent with a saddle-node-type bifurcation. Beyond the periodic regime, we find spatiotemporal chaos at large applied pressure gradients. The spatiotemporal chaos organizes itself via continuous creation, annihilation and advection of defects along the channel in an irregular manner.


Our findings highlight the need for systematic experimental investigations of confined channel flows in nematic fluids. In the absence of external forcing, equilibrium states exhibiting spatially varying order parameters $\mathbf{Q}$ and boundary-layer structures in the nematic order should be characterized. The multistability observed under applied pressure gradients could be explored through hysteresis experiments, wherein the imposed pressure gradient is varied gradually to look for the coexistence of and switching between Bowser and Dowser states. At higher flow rates, where our analysis predicts oscillatory dynamics and defect-mediated transport, experiments could seek signatures of periodic director oscillations and advecting defect states. Collectively, such measurements would provide direct experimental validation of the dynamical regimes identified here, bridging theoretical predictions and observable behaviour in nematic channel flows.


A useful comparison may be drawn between the spatiotemporal chaos reported here and the turbulent regimes of active nematics. Active nematic turbulence is generally regarded as a two-dimensional instability, sustained by self-propelled $\pm 1/2$ defects. By contrast, the defect-mediated turbulence we report here arises even within our one-dimensional reduction of the Beris–Edwards model, where it manifests through oscillatory instabilities and chaotic dynamics of the order parameter and director field. As a result, the chaotic regimes we observe can already arise in 1D, with 2D simulations enriching this picture through advecting defects. This highlights a key distinction: whereas active nematic turbulence is characterized by two-dimensional defect dynamics, the spatiotemporal chaos of pressure-driven nematics is rooted in confinement-induced instabilities that persist even in reduced-dimensional descriptions.

It will be interesting to investigate how these instabilities are modified in three-dimensional geometries and under different confinement conditions. Such extensions would clarify the extent to which the mechanisms identified here are generic features of confined nematic flows. Beyond passive systems, an extension of our analysis to active nematics is especially promising. In this case, the equilibrium state with spatially varying $\mathbf{Q}$ would generate thresholdless active flows due to the presence of director gradients. The competing effects of active turbulence and pressure-driven turbulence may give rise to a rich dynamical landscape, which we leave for future work.



\section*{Acknowledgements} 
R.V. acknowledges the support of the Leverhulme Trust [Grant No. LIP-2020-014]. R.V. and J.M.Y. acknowledge the support of the ERC Advanced Grant ActBio (funded as UKRI Frontier Research Grant EP/Y033981/1). S.P.T. thanks the Royal Society and the Wolfson Foundation for the Royal Society Wolfson Fellowship award and acknowledges the support of the Department of Science and Technology, India via the research grant CRG/2023/000169. 

\section*{Declaration of Interests}
The authors report no conflict of interest.

 \appendix

\section{One-dimensional nematohydrodynamics equations}\label{App: oneD}

Here we present the reduced {nematohydrodynamic equations} for the case of a unidirectional flow. For unidirectional flows i.e. a $1$D formalism, we consider variations only in the direction across the width of the channel. This results in the following nematohydrodynamic equations:


\begin{align}\label{Eq: 1D Qvx varyS}
\frac{\partial Q_{xx}}{\partial t} &= \Gamma K \frac{\partial^2 Q_{xx}}{\partial y^2} 
+ Q_{xy} \frac{\partial u_x}{\partial y} 
+ \Gamma C Q_{xx} \left(S^2_{\mathrm{nem}} - S^2\right) 
- 2 \lambda Q_{xx} Q_{xy} \frac{\partial u_x}{\partial y}, \\ \nonumber
\frac{\partial Q_{xy}}{\partial t} &= \Gamma K \frac{\partial^2 Q_{xy}}{\partial y^2} 
- Q_{xx} \frac{\partial u_x}{\partial y} 
+ \Gamma C Q_{xy} \left(S^2_{\mathrm{nem}} - S^2\right) 
+ \frac{\lambda}{2} \frac{\partial u_x}{\partial y} 
- 2 \lambda Q_{xy}^2 \frac{\partial u_x}{\partial y}, \\ \nonumber
\rho \frac{\partial u_x}{\partial t} &= \eta \frac{\partial^2 u_x}{\partial y^2} 
+ F + F_{\text{b}}.
\end{align}

The backflow contribution is given by:
\begin{align}
F_{\text{b}} = \frac{\partial \sigma_{xy}}{\partial y} &= 2 \Big( H_{xy} \partial_y Q_{xx} + Q_{xx} \partial_y H_{xy}-H_{xx} \partial_y Q_{xy} - Q_{xy} \partial_y H_{xx} \Big) \\ \nonumber
& + \lambda \Big( 4 Q_{xx} Q_{xy} \partial_y H_{xx} + 4 H_{xx} Q_{xy} \partial_y Q_{xx} + 4 H_{xx} Q_{xx} \partial_y Q_{xy} \\ \nonumber
& \quad + 4 Q_{xy}^2 \partial_y H_{xy} + 8 H_{xy} Q_{xy} \partial_y Q_{xy} - \partial_y H_{xy} \Big),
\end{align}


where the scalar order parameter is $S^2 = Q_{xx}^2 + Q_{xy}^2$, and the molecular fields are
\begin{align}
H_{xx} &= C Q_{xx} \left(S_{\mathrm{nem}}^2 - S^2\right) + K \frac{\partial^2 Q_{xx}}{\partial y^2}, \\ 
H_{xy} &= C Q_{xy} \left(S_{\mathrm{nem}}^2 - S^2\right) + K \frac{\partial^2 Q_{xy}}{\partial y^2}.
\end{align}
Here, the terms in the $Q$-tensor evolution equations represent: elastic relaxation of distortions ($\Gamma K \partial^2_y Q_{ij}$), flow alignment ($Q_{ij} \partial_y u_x$ and related $\lambda$ terms), and relaxation towards the nematic order parameter ($\Gamma C Q_{ij}(S_{\mathrm{nem}}^2 - S^2)$). In the momentum equation, viscosity ($\eta \partial_y^2 u_x$), external forcing ($F$), and nematic backflow effects ($F_{\text{b}}$) are included. The backflow term accounts for the feedback of director reorientation on the fluid velocity through the coupling between the $Q$-tensor and the molecular field {tensor $H$}.


\section{Boundary layer analysis for varying $\mathbf{Q}$ state}\label{App: boundary layer}

In this section, we {perform} a boundary layer analysis for the varying $\mathbf{Q}$ state for the boundary layers that are formed for $S$ near the walls for small $\epsilon$. Considering the steady state{,} Eqs.~\eqref{Eq: 1D Sthetavx}, 
the second equation {can be rewritten} as
$$\frac{d}{dy}\left(S^2(y)\frac{d\theta}{dy}\right)=0,$$
giving
$$S^2(y)\frac{d\theta}{dy}=m,$$
where $m$ is a constant. Substituting this in the first equation of Eqs.~\eqref{Eq: 1D Sthetavx} for $S$ we obtain
\begin{equation}\label{Eq: S eq}
 \frac{d^2 S}{dy^2} - \frac{4 m^2}{S^3}+\frac{C}{K}S(1-S^2)=0.   
\end{equation}
The value of $m$ is determined as $m=\theta'(\pm L/2)$ since $S(\pm L/2)=1$. Hence for {the} uniform $\mathbf{Q}$ state we have $m=0$ and for  {the} lowest energy ($n=1$) varying $\mathbf{Q}$ state we have $m\approx\pm\pi/L$. 

Substituting $m\approx\pm\pi/L$ in equation~\eqref{Eq: S eq} and non-dimensionalizing with length scale $L/2$, {gives} the following ODE:
\begin{equation}\label{eq: S_nondim}
  4\epsilon  \frac{d^2 S}{dy^2}  - \frac{4\pi^2 \epsilon}{S^3}  +S(1-S^2)=0.
\end{equation}
Here $\epsilon=K/(CL^2)$ and we assume $0<\epsilon\ll1$ with the boundary conditions of the ODE $S(-1)=S(1)=1$.

\subsection{Outer solution in the channel bulk}

Setting $\epsilon=0$, we get the equation
$$S(1-S^2)=0,$$
implying $S^{(0)}_{out}=1$ as the physically relevant solution. Performing a regular perturbation expansion of the outer solution {gives}
$$S_{out}(y)=1+\epsilon S^{(1)}_{out}+O(\epsilon^2).$$
Substituting
and balancing terms at $O(\epsilon)$ {leads to}
$$S^{(1)}_{out}=-2\pi^2.$$ Hence, the outer (bulk) solution to $O(\epsilon)$ is
$$S_{out}(y)=1-2\pi^2 \epsilon + O(\epsilon^2).$$

\subsection{Inner solution near the channel boundaries}

We consider the top boundary without loss of generality, and define a rescaled variable,
$$Y=\frac{1-y}{\delta},$$
where $0<\delta\ll1$. We now write the full solution to $O(\epsilon)$ as
$$S(y)=1-2\pi^2 \epsilon+v(Y),$$
where $v(Y)\sim O(\epsilon)$.
Substituting the expansion in equation~\eqref{eq: S_nondim} {gives}
\begin{align*}
4\epsilon\frac{1}{\delta^2}\frac{d^2 v}{d Y^2}-4\pi^2\epsilon\frac{1}{(1-2\pi^2 \epsilon+v)^3}
+(1-2\pi^2 \epsilon+v)(1-(1-2\pi^2 \epsilon+v)^2)=0.
\end{align*}
Substituting $\delta=\sqrt{2\epsilon}$ and keeping terms of ${O}(1)$ we get,
$$v(Y)=B e^{-Y},\,\,Y\geq 0.$$
This solution decays in the bulk. To find $B$, we use the boundary condition $S(1)=1$ giving $v(0)=B=2\pi^2 \epsilon$ and {leading to a solution}
$$v(y)=2\pi^2 \epsilon\,e^{-\frac{1-y}{\sqrt{2\epsilon}}}.$$

A similar boundary layer appears near the bottom wall $y=-1$ with $Y=(y+1)/\delta$.
Hence, the total solution of the system is given by
$$S(y)=1-2\pi^2\epsilon + 2\pi^2\epsilon \left( e^{-\frac{1-y}{\sqrt{2\epsilon}}} + e^{-\frac{1+y}{\sqrt{2\epsilon}}} \right) + O(\epsilon^2).$$


\section{Additional results for different parameter combinations}\label{App: 1}

\begin{figure}
\centering
\captionsetup{width=\columnwidth}
\includegraphics[width=0.9\columnwidth]{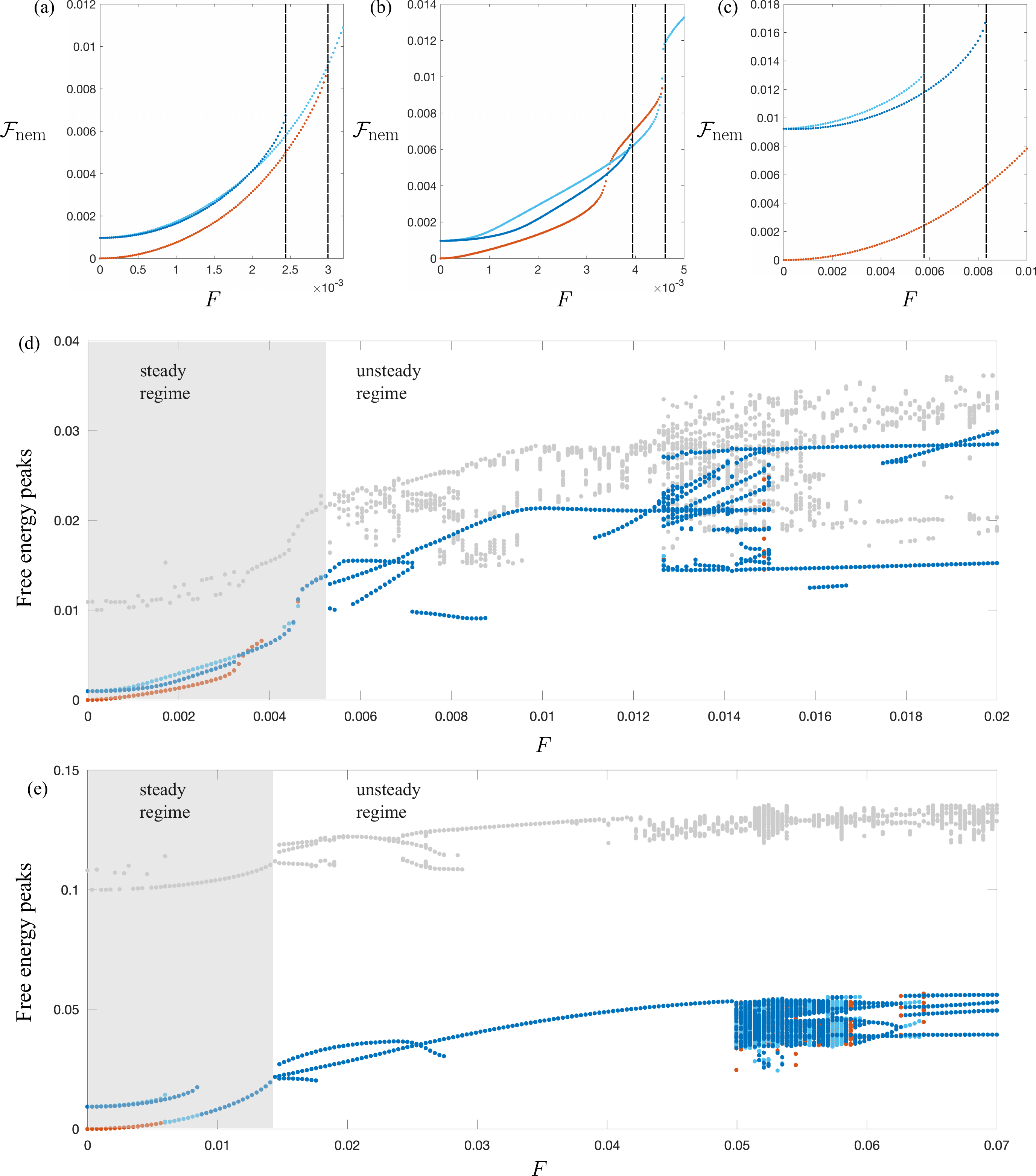}
\caption{Steady flow states for pressure-driven nematics in a channel for parameters (a) $C=0.025$, $K=0.01$, $L=20$, $\lambda=0.2$, (b) $C=0.025$, $K=0.01$, $L=20$, $\lambda=1$ and (c) $C=0.025$, $K=0.1$, $L=20$, $\lambda=0$. Vertical dashed lines show when the solution branches become unstable. (d) and (e) show the full bifurcation diagram where the peaks in the time series of $\mathcal{F}_{\mathrm{nem}}$ are plotted as a function of $F$ for the parameters corresponding to (b) and (c), respectively. In these panels, initial flow velocity is zero and different colours correspond to different initial orientation profiles with red favouring Bowser states, blue and cyan favouring Dowser states and gray corresponding to random initial orientations.}
\label{Fig: A1}
\end{figure}

Here we {present} further plots showing the free energy changes of {the} Bowser and Dowser states as a function of {the forcing} $F$ for {different} parameter values. For a non-zero flow aligning parameter, the plot in figure~\ref{Fig: 2_1}(a) {is} modified to figure~\ref{Fig: A1}(a) ($\lambda=0.2$) and figure~\ref{Fig: A1}(b) ($\lambda=1$). {Keeping} $\lambda=0$ and {increasing} $K$, {gives} the plot in figure~\ref{Fig: A1}(c) where we see that the {ordering of the instabilities} has changed, i.e.~the two Dowser states become unstable before the Bowser state. For the parameter values corresponding to (b) and (c), a full bifurcation diagram {for} the $1$D model is shown in figure~\ref{Fig: A1}(d) and (e), respectively. In figure~\ref{Fig: A1}(d), we still observe the unsteady periodic and chaotic behaviours in the flow aligning regime corresonding to $\lambda=1$. In figure~\ref{Fig: A1}(e), which corresponds to a larger elasticity $K$ we again still observe unsteady flows. Furtheremore, here we find that the steady states are stable for larger $F$ values compared to figure~\ref{Fig: 3}(a), as well as an extended window of periodic behaviours in the unsteady regime.

\bibliographystyle{jfm}

\bibliography{nematic_channel}

\end{document}